\documentclass[prl,twocolumn,showpacs,floatfix]{revtex4}
\usepackage{graphicx}
\usepackage{amsmath}
\usepackage{bm}
\usepackage{braket}
\DeclareGraphicsExtensions{.jpg,.png,.pdf}

\begin{document}
\title{Electronic transport in BN-substituted bilayer graphene nano-junctions}
\author{Daniele Giofr\'e}
\affiliation{Dipartimento di Scienze dei Materiali, Universit\`a di
Milano Bicocca, via Cozzi 53, 20126 Milan, Italy.}

\author{Davide Ceresoli}
\affiliation{Istituto di Scienze e Tecnologie Molecolari (CNR-ISTM) CNR,
via Golgi 19, 20133 Milan, Italy}

\author{Mario I. Trioni}
\thanks{Corresponding author}
\email{mario.trioni@istm.cnr.it}
\affiliation{Istituto di Scienze e Tecnologie Molecolari (CNR-ISTM) CNR,
via Golgi 19, 20133 Milan, Italy}

\date{\today}

\begin{abstract}
We investigated a suspended bilayer graphene where the bottom (top)
layer is doped by boron (nitrogen) substitutional atoms by using Density
Functional Theory (DFT) calculations. We found that at high dopant
concentration (one B-N pair every 32 C atoms) the electronic structure
of the bilayer does not depend on the B-N distance but on the relative
occupation of the bilayer graphene sub-lattices by B and N.  We found that
a large built in electric field is established between layers, giving
rise to an energy gap. We further investigated the transport properties
and found that intra-layer electron current is weakly influenced by the
presence of these dopants while the inter-layer current is significantly
enhanced for biases allowing the energy alignment of N and B states. This
effect leads to current rectification in asymmetric junctions.
\end{abstract}

\maketitle

Bilayer graphene (BLG) is a two-dimensional material constituted by two
stacked graphene layers. BLG has recently attracted much interest because
it show exceptionally high charge mobility like in single layer graphene.
Charge transport in BLG is tied the in-plane direction of each of the
two layers. As a consequence, BLG has been proposed to be more suitable
than single layer graphene to realize carbon-based, field effect logic
devices~\cite{Fiori2009, Bane2009,li11}. 
However, due to the small density of states of BLG
at the Fermi level, in order to apply a gate voltage across BLG, very
large electric fields are needed.  This has deleterious impact on the
performance and stability of gated BLG. For instance, graphene and BLG
deposited on oxide dielectric such as SiO$_2$ show a large hysteresis
in the I-V curve, due to the charging/discharging of point defects in
the oxide. Moreover, charged impurities in the gate oxide and
at the graphene-gate interface,have been shown to reduce dramatically
the mobility of graphene-based devices~\cite{lafkioti10,liao10}.

On the other hand, the low density of states can be exploited to induce
a large built it electric field in BLG. In fact, even the addition or
removal of a small amount of charge from each layer causes an abrupt
change of the Fermi level. In this paper we simulate a suspended bilayer
graphene nano-junctions where boron and nitrogen atoms substitute
carbon atoms from the top and bottom layers, respectively. The equal
number of B and N atoms assures that the system is neutral. By first
principles calculations we have found that a large built in electric
field is then established, which is the consequence of two effects: the
first is the Fermi-level mismatch between the B-rich and N-rich layer,
the second is a partial charge transfer between the two layers. As
a consequence, a small energy gap appears enabling this system to be
employed for nano-electronics applications. In view of this, we also
investigated the ballistic transport in a nano-junction containing B
and N substitutional atoms.

We performed DFT calculations with the local-basis SIESTA
code~\cite{SIESTA1999,30Artacho,29Soler,28Ordejon}.  We used
norm-conserving pseudopotentials~\cite{32Troullier} and the DZP
basis set. The exchange correlation potential is described by the
PBE functional~\cite{26Perdew} and the weak van der Waals forces
between the graphene layers are described by a classical $C/r^6$ term,
parametrized by Grimme~\cite{Grimme2004, Grimme2006}.  With this method,
we obtained an interlayer distance of 3.32 \AA\ for pristine BLG,
in good agreement with the experimental value of 3.36 \AA~\cite{}.
In every subsequent calculation, we fixed the in-plane lattice
constant to the calculated value is 2.474 \AA, which is close to the
experimental value of 2.46 \AA~\cite{}. We employed the slab-dipole
correction~\cite{SIESTA1999}. 

We simulated the doping of each layer by constructing a 4$\times$4
supercell i.e. 32 carbon atoms per layer with a single B-N substitutional
pair.  The electronic properties of the system depends on the the
relative position of the two dopants in the graphene sub-lattices.
In fact, graphene is a bipartite lattice and the for BLG the two sub
lattices are no more equivalent. The first sub-lattice (S1) is constituted
by carbon atoms sitting on top of carbon atoms from the other layer.
The second sub-lattice (S2) is constituted by carbon atoms positioned
above the geometrical center of the hexagons from the other layers.

\begin{figure}\begin{center}
  \includegraphics[width=0.9\columnwidth]{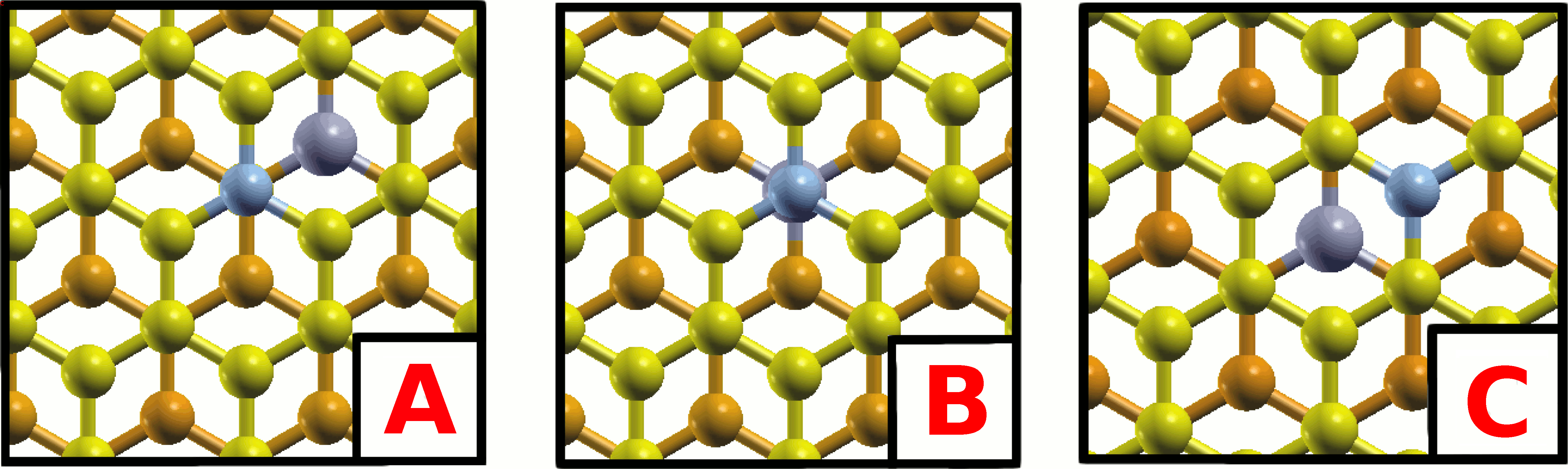}
  \caption{Top view of the possible arrangement model geometries for a BN
  pair substitutional in graphene bilayer. Grey sphere: B; blue sphere: N;
  yellow/brown sphere are C atoms of the two inequivalent sublattices.}
  \label{fig:1}
\end{center}\end{figure}

Thus, in BN-substituted BLG, with only one BN pair, only three
configurations are permitted. The first is when B and N belong to
different sub-lattices (\textbf{A} in Fig.~\ref{fig:1}); the second is
characterized by B and N belonging to the first sub-lattice and are one
on top of each other (\textbf{B} in Fig.~\ref{fig:1}).  In the third
configuration B and N occupy both the S2 sub-lattice (\textbf{C} in
Fig.~\ref{fig:1}).  In the 4$\times$4 supercell, additional arrangements
are possible which differ from \textbf{A}, \textbf{B} and \textbf{C} for
the relative B-N distance.  We verified that for a given BN-substitution
of the two sublattices, the distance between the impurities has a minor
influence the electronic properties of the system. This is reasonable
because the added/removed electron in each graphene layer is almost
completely delocalized in its own sub-lattice.

In the rest, we concentrate on the most energetically stable system,
i.e. the configuration \textbf{C}. This configuration is $\sim$~0.06~eV
lower in energy than \textbf{B} and $\sim$~0.07 eV than \textbf{A}.
Configuration \textbf{C} is insulating, while \textbf{A} and \textbf{A}
are metallic.

\begin{figure}\begin{center}
  \includegraphics[width=0.9\columnwidth]{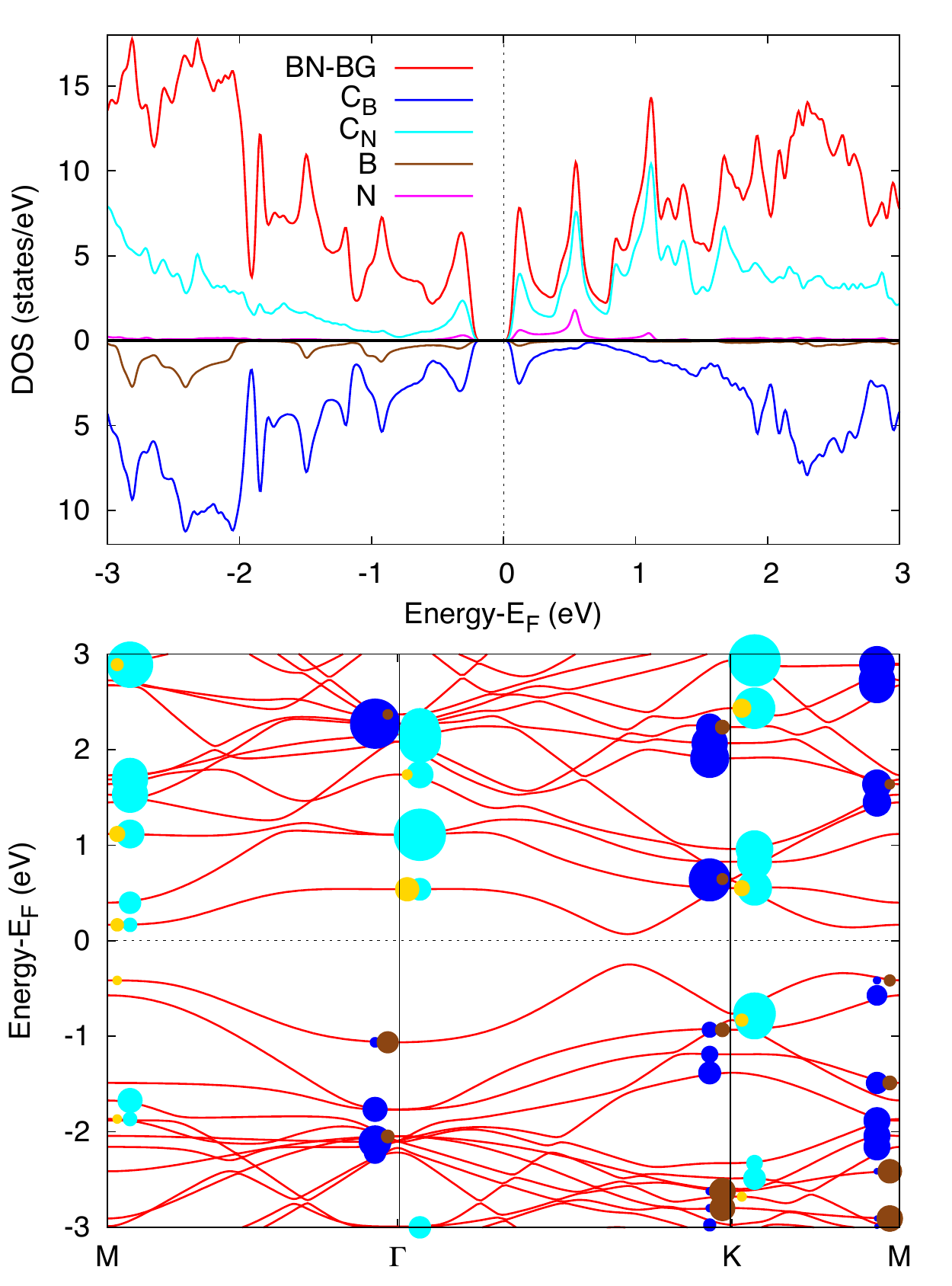}
  \caption{Top panel: Density of states of the BN doped BLG. The projected
  DOSs on N and carbon atoms of the N doped graphene layer are reported
  in the upper part, while those projected on B and carbon atoms of the
  B doped graphene in the lower part.  Bottom panel: band structure
  of BN doped BLG. The size of the circles represent the weight of
  the eigenstates on specific atoms according to the following color
  code: yellow: N , light blue: C in N-doped layer, brown: B, blue:
  C in B-layer.}
  \label{fig:2}
\end{center}\end{figure}

In Fig.~\ref{fig:2} we report the density of states (DOS) and the band
structure of this configuration. The system is an insulator, with an
indirect Kohn-Sham band gap of 0.35~eV, between the high symmetry points K
and M.  Moreover, the characteristic ``V-shaped'' DOS of graphene centered
around the Dirac point, is still visible in the plot, except that the two
linear branches are separated away by an energy $U$~=~1.9~eV.  The DOS
of the system is similar to that of pristine BLG but with an external
applied electric field of 0.37~V/\AA\ along the direction perpendicular
to the sheets.
 
To rationalize the energy separation $U$ between the Dirac, we performed
additional calculations of the two isolated doped layers and of a
pristine graphene layers, and of BLG in presence of perpendicular
electric field. We have found that $U$ is given by the sum of two
contributions: $U = U_{CT} + U_\mathrm{EF}$ where the first term is
the built in electric field generated by the Fermi-level mismatch, in
absence of charge transfer (i.e. when the two layers are separated by a
large distance) and the second term is due to charge transfer between the
two layers.  Thus, the first term represents a purely classical capacitor,
whereas the second term is of purely quantum origin, and depends on the
overlap of the wavefunctions belonging to the two different layers. A
full account of the derivation of $U_{EF}$ and $U_{CT}$ is presented in
the additional material.

From the band structure one can also note that the highest occupied band
belongs to the B-doped layer while the lowest unoccupied belongs to the
N-doped layer.  The difference of the electronic structure between the two
graphene layers with donor or acceptor dopants, motivated us to simulate
the electronic ballistic transport through BN-doped BLG nano-junctions.

The first principles quantum transport calculations were performed using
the TranSIESTA code~\cite{38Brandbyge}, which employs the non-equilibrium
Green's function (NEGF) formalism in conjunction with DFT.  Within the
NEGF-DFT formalism, the system has been divided into three regions: the
left and right electrodes, and the transition region.  The transition
part contains the portion of physical electrodes, the so-called right
and left junction, where all screening effects take place. In order to
apply an external bias, the Fermi level of the electrodes are shifted
relative to each other and the electronic occupations of the system are
determined by electrochemical potentials of electrodes.

\begin{figure}\begin{center}
  \includegraphics[width=0.9\columnwidth]{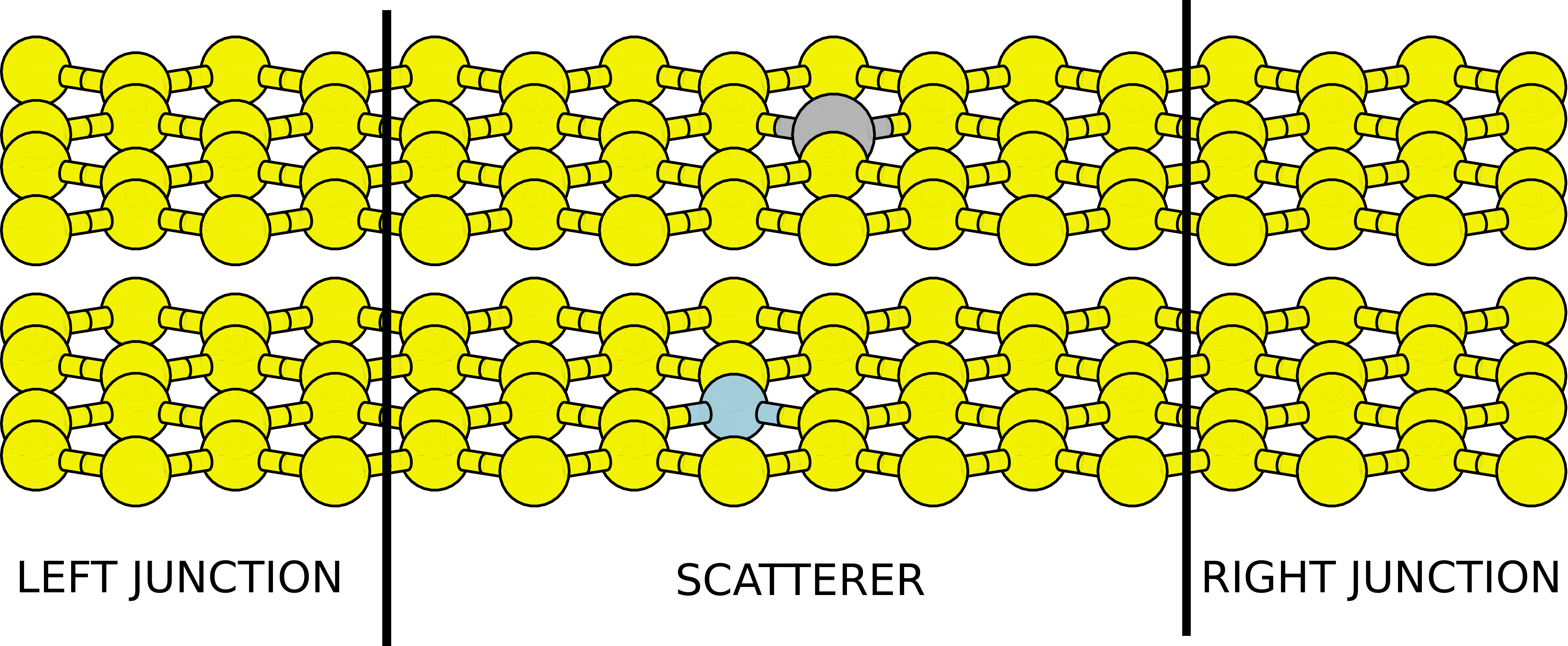}
  \caption{Transport geometry of the ``overlay'' BN-doped junction. The
  upper layer contains the nitrogen atom while the lower layer contains
  the boron one.}
  \label{fig:3}
\end{center}\end{figure}

We first consider the electron transport through a finite BN-doped
region.  The left and right electrodes are made of pristine BLG while
the scattering region containing a single pair of BN-doping BLG, as
shown in Fig.~\ref{fig:3}. The scattering region contains 64 atoms,
while the electrodes are made of 32 atoms each.

\begin{figure}\begin{center}
  \includegraphics[width=0.9\columnwidth]{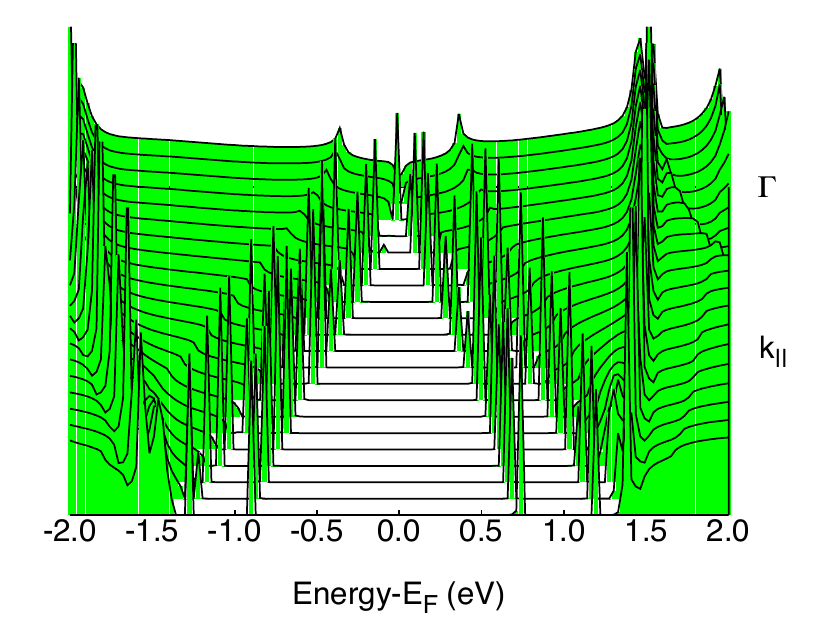}
  \caption{$k_{\parallel}$-resolved DOS of the ``overlay'' nano-junction
  shown in Fig.~\ref{fig:3}.}
  \label{fig:4}
\end{center}\end{figure}

Although we have found that the two dopants represent a weak perturbation
to BLG, they can nevertheless play a role in transport, by reducing
the number of available channels.  The $k_{\parallel}$-resolved DOS
of the system is shown in Fig.~\ref{fig:4} together with those of
prisitne BLG.  The effect of doping is the appearance of new states:
at about $-$2~eV at $\overline\Gamma$ ($k_{\parallel}=0$) the occupied
B~$p_z$ orbital, with a small dispersion on $k_{\parallel}$.  The state
about $+$1.5~eV is the N~$p_z$ nitrogen orbital, showing a negligible
dispersion along $k_{\parallel}$.

Additionally, four other states appear in the projected-DOS gap. Two
states which are very close the band edges behave like Tamm-like
states~\cite{Tamm1932} emerging from the graphene bands. The one at
negative energy has a large boron character and is localized on the
B-doped graphene layer while that at positive energy is the N-doped
layer counterpart. The remaining pair of states are Shockley-like
states~\cite{Shockley1939} are are due to the discontinuity in the
potential felt by the electron, in strict analogy with the classical
surface states. Both these Tamn and Schockley are 1-D in character and
are strongly localized. For this reason, they do not contribute to the
electronic transport but they behave as a scattering centers.

We calculated the $I(V)$ curve with the NEGF-DFT method, where the charge
density is self-consistently updated upon the effect of the applied bias.
Once the transmission probability is worked out, the current is obtained
integrating in the energy window corresponding to the applied voltage.

\begin{figure}\begin{center}
  \includegraphics[width=0.9\columnwidth]{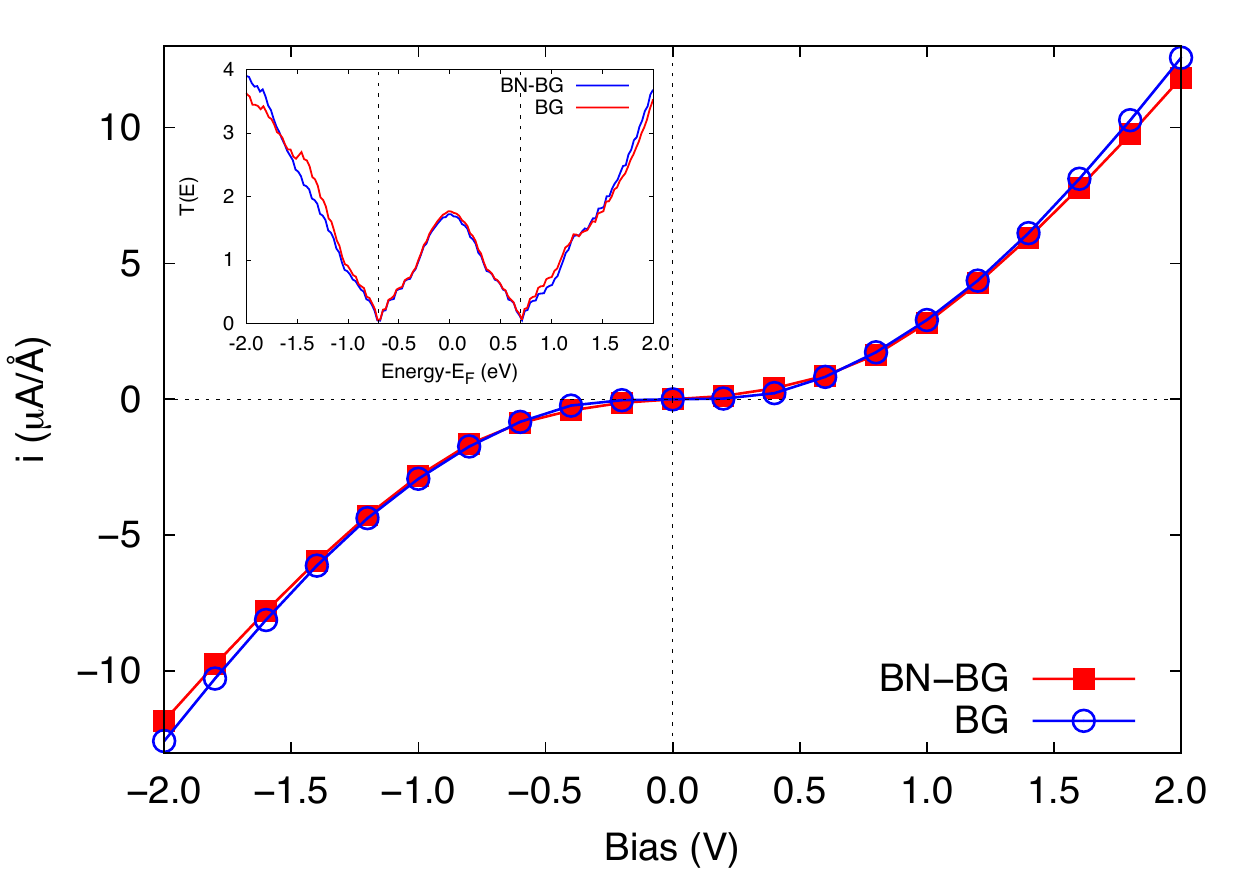}
  \caption{$I(V)$ characteristic of the ``overlay'' nano-junction shown in
  Fig.~\ref{fig:4}. The inset reports the transmission probability at a
  bias potential of $-$1.4~eV.}
  \label{fig:5}
\end{center}\end{figure}

We show in Fig.~\ref{fig:5} the $I(V)$ curve for the doped BLG compared
to the pristine BLG.  The comparison shows that the I-V characteristics
are very similar and that the current flowing in the BN-doped system is
slightly smaller. Indeed the N and B states do not perturb the system in
the energy range relevant for transport and the C network is preserved.
Furthermore, the inter-layer current flow is much smaller than the
intra-layer one.  As an example, in the inset of Fig.~\ref{fig:5}
is reported the transmission probability for these two systems at an
applied voltage of $-1.4$~eV. They look very similar, especially in
the bias energy window, further confirming our expectation.

\begin{figure}\begin{center}
  \includegraphics[width=0.9\columnwidth]{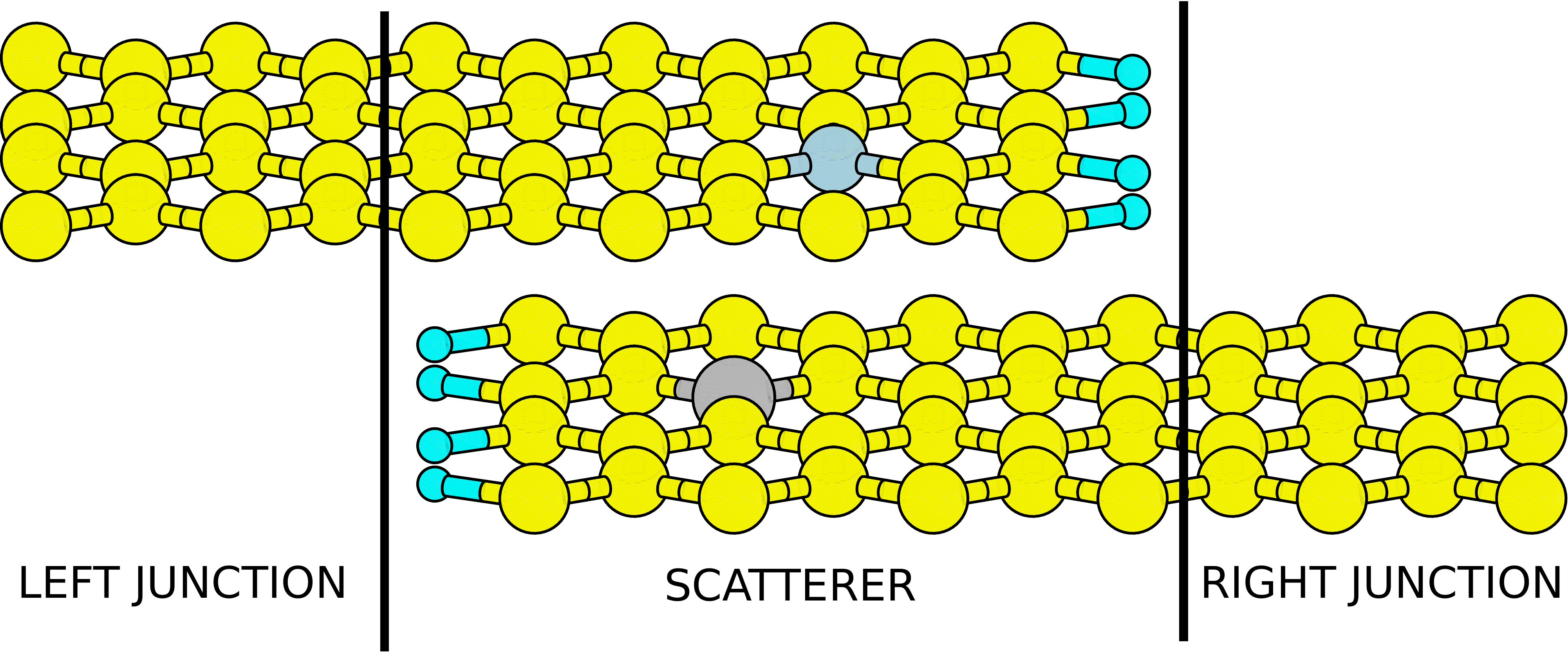}
  \caption{Geometry of the  BN-doped ``shingle'' nano-junction. The
  upper layer contains the nitrogen atom while the lower layer contains
  the boron one.}
  \label{fig:6}
\end{center}\end{figure}

We now consider a ``shingle'' nano-junction, constituted by two semi
infinite H-terminated graphene layers where the dopants are located in
the contact region (Fig.~\ref{fig:6}).  The electronic transport across
this junction is a very interesting case, showing that when the carbon
network is broken, for example when the graphene bilayer is formed
by flakes overlapping to some extend, rectification can occur.  In the
present case, the electron current must flow from one layer to the other,
overcoming an energy barrier this system breaks the left-right symmetry
and behaves like a n-p nano-junction.

By comparing this nano-junction with the same system without BN-dopants,
it is possible to study the effect of the single BN couple on the
transport properties. As in the previous case the  junction is completely
contained in the scattering region, but the electrodes are now constituted
by semi-infinite graphene layers.

\begin{figure}\begin{center}
  \includegraphics[width=0.9\columnwidth]{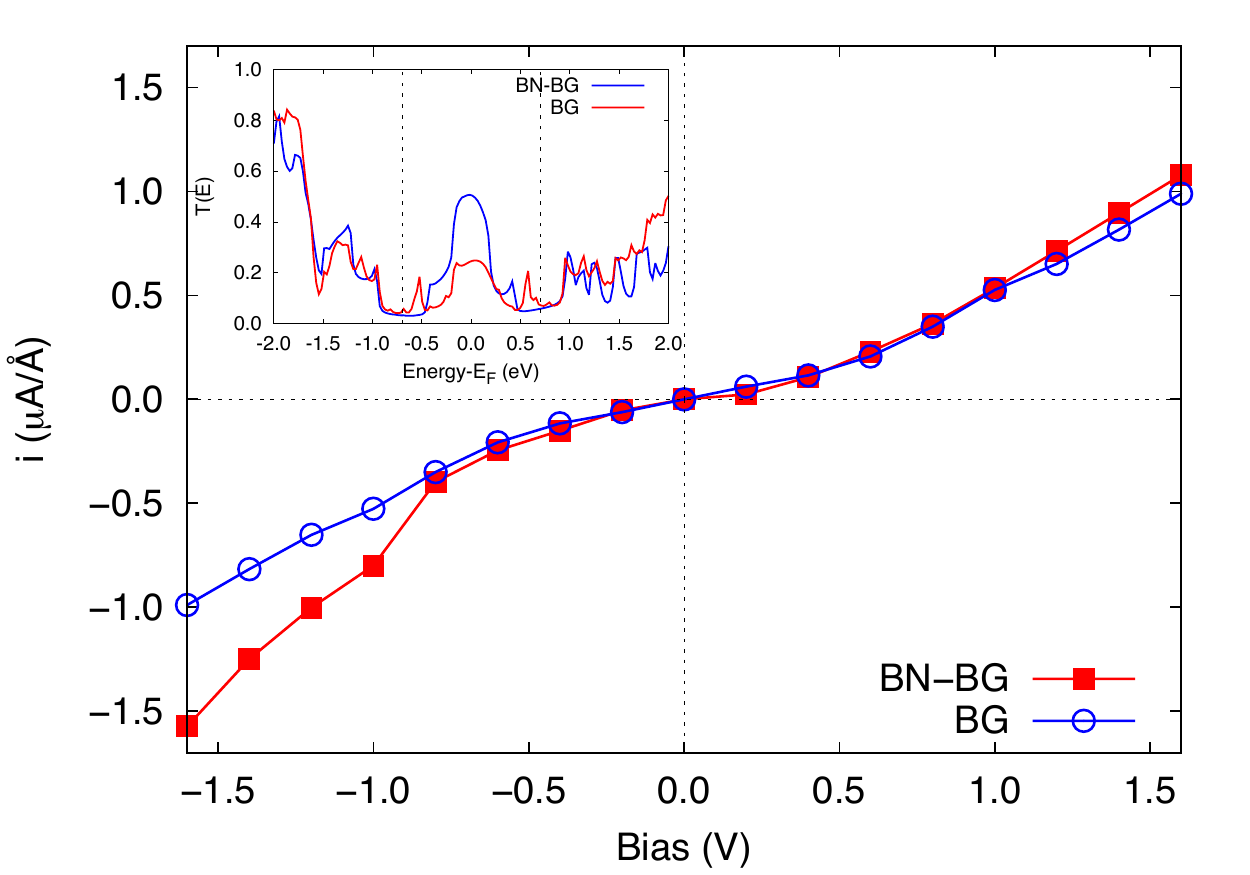}
  \caption{$I(V)$ characteristic of the ``shingle'' nano-junction shown in
  Fig.~\ref{fig:6}. The inset reports the transmission probability at a
  bias potential of $-$1.4~eV.}
  \label{fig:7}
\end{center}\end{figure}

The computed I-V characteristic is reported in Fig.~\ref{fig:7}. The
first finding is the reduction of the absolute value of the current by
one order of magnitude. This reduction does not suppress completely
the current flow because the partial overlap of the $\pi$ bands of
graphene on different layers.  We also note that for positive bias
(i.e. for electrons which flow from the N-doped to the B-doped layer)
the electron current through the doped junction is a bit larger then
those obtained without the BN-pair. Therefore, the presence of these
impurities play the role to lower the potential barrier between layers.

The most interesting results is the behavior of the junction at
large negative biases: in this case the current of the doped system
is significantly larger (of about 50\%) than the pristine one.
By inspecting the transmission functions (at $-$1.4~V bias, see the
inset of Fig.~\ref{fig:7}), we found that this increase is not due to
the entry of N or B related features in the energy window relevant for
transport but instead to an increase of the tunnel probability in the
middle of this range.  A further analysis of the DOS confirms that this
effect is due to the alignment of the N and B energy resonances, now
shifted by the bias.  In fact, for bias lower than $-$1.0~V, the filled
B state and the empty N state start to align, allowing a preferential
channel for interlayer current.

In conclusion, we studied the electronic and transport properties of a
N and B doped graphene bilayer by first principles and we found that a
strong electric field is established between the two layer because the
opposite nature of the dopants (acceptor and donor), also creating a
small energy gap.  Our analysis showed a small effect of the dopants
on the transport properties of the graphene bilayer.  Differently,
when the carbon network is broken like for overlapping graphene flakes,
the interlayer current start to play the major role. In this case the
presence of the dopants enhances the current but for negative biases only,
originating a rectification effect in this n-p nano-junction.



\end{document}